\documentstyle[12pt,epsfig]{article}

\def \to {\rightarrow}
\def \beq {\begin{equation}}
\def \eeq {\end{equation}}
\def \ba {\begin{eqnarray}}
\def \ea {\end{eqnarray}}
\def \jpsi {J/\psi}
\def \mtss {<\!0|{\cal O } ^{J/\psi}_{1} [ ^3S_1 ]|0\!> }
\def \mtso {<\!0|{\cal O } ^{J/\psi}_{8} [ ^3S_1 ]|0\!> }
\def \moso {<\!0|{\cal O } ^{\jpsi}_{8} [ ^1S_0 ]|\!0> }
\def \mtpo {<\!0|{\cal O } ^{\jpsi}_{8} [ ^3P_0 ]|\!0> }
\def \mtpj {<\!0|{\cal O } ^{\jpsi}_{8} [ ^3P_J ]|\!0> }

\begin{document}
\title{Associated $ \jpsi + \gamma $ diffractive production:  the nature
of Pomeron and test of hard diffractive factorization}
\author{Jia-Sheng Xu \\
{\small {\it Department of Physics, Peking University,
Beijing 100871,  China }} \\
Hong-An Peng \\
{\small {\it China Center of Advance Science and
Technology (World Laboratory), Beijing 100080, China }} \\
{\small {\it and Department of Physics, Peking University, Beijing 100871,
 China} }
}
\date{}
\maketitle
\begin{abstract}
We present a study of diffractive associated $\jpsi + \gamma$ production
at the Fermilab Tevatron and LHC based on the Ingelman-Schlein model
for hard diffractive scattering and the factorization formalism of NRQCD
for quarkonia production. we find that this process $(p + {\bar p} \to
p + \jpsi + \gamma + X)$ can be used to probe the gluon content of the
Pomeron and test the assumption of diffractive hard scattering
factorization. Using the renormalized Pomeron flux factor $D\simeq 0.11
(0.052)$ , the single diffractive $ \jpsi + \gamma $ production cross
section at $4 < P_T < 10 $ GeV , $ -1 < \eta < 1 $ region is found to
be of the order of $ 3.0 $pb ($ 8.5 $ pb ). The ratio of single
diffractive to inclusive  production is
$0.50 \% (0.15 \%) $ in central region at the Tevatron (LHC) for the
gluon fraction in the Pomeron $f_g = 0.7 $, independent of the values
of color-octet matrix elements.
\vskip 3mm
\noindent PACS number(s): 12.40Nn, 13.85.Ni, 14.40.Gx
\end{abstract}

\vfill\eject\pagestyle{plain}\setcounter{page}{1}

\section{Introduction}

Early in the 1960's physicists already well realized that in high energy
strong interaction the Regge trajectory with vacuum quantum number, the
Pomeron, plays a particular and very important role in soft
processes, such as the energy dependence of $ \sigma_{tot} (s)$, the
behavior of the elastic differential cross section $ \frac{d \sigma^{el} }
{dt} $at small $|t|$, the single and double diffractive dissociation
processes in hadron-hadron collisions \cite{collins}. After the advent of
QCD , physicists begin to study the nature of the pomeron in the QCD
framework \cite{low}. Since the Pomeron carries the quantum number of
the vacuum, in QCD language, it is a colorless entity, which leads to
the diffractive events are characterized by a large rapidity gap, a
region in rapidity devoiding of hadronic energy flow, this distinct class of
events are observed by the ZEUS and H1 Collaboration at HERA in deep
inelastic scattering region \cite{gap}, which offer a unique chance to
study the the soft scattering process with a hard virtual photon probe,
therefore provide a more complete understanding of QCD .

\par
In an orginal paper \cite{ingelmam}, It was suggested that hard diffractive
scattering processes would give new and valuable insight about the nature
of Pomeron. the Ingelman-Schlein (I-S) model assumes that the hard
diffractive scattering processes can be calculated in a factorized way :
first, a Pomeron is emitted from the diffractively scattered hadron, then
partons of the Pomeron take part in hard subprocesses, the results will
be diffractively produced high-$P_T$ jets in $P - P({\bar P}), \gamma - P$
collisions or larger rapidity gap events in diffractive deep inelastic
scattering (DDIS) , so the partonic struction of the Pomeron could be
established and studied experimentally. These pridictions were confirmed
subsequently. The UA8 Collaboration at CERN S $p{\bar p}$S collider with $
\sqrt{s} = 630$ GeV studied the diffractive jet distribution , indicates a
dominant hard partonic structure of the Pomeron, however, the diffractive
dijet event topology alone can not distinguish between a hard-quark  or a
hard-gluon struction in the Pomeron \cite{ua8}.
The DDIS and dijet photoproduction experiments at HERA have
shed light on the partonic structure of the Pomeron. Combining the
measurements of the diffractive structure function in DDIS and the
photoproduction jet cross sections, the ZEUS Collaboration gives the first
experimental evidence for the gluon content of the Pomeron and determines
that the hard gluon fraction of the pomeron $f_g$ is in the range
$ 0.3 < f_g < 0.8 $, independent of the validity of the momentum sum rule
for the Pomeron and the normalization of the flux of Pomeron from the
proton\cite{zeus} .  The H1 Collaboration also determines the fraction of
the momentum of the Pomeron carried by the hard gluon, which is
$ f_g \sim 0.9 $ at $ Q^2 = 4.5 $ GeV$^2$ and $ f_g \sim 0.8 $ at $Q^2 = 75$
GeV$^2 $ \cite{h1} . The partonic structure of the Pomeron, based on the
I-S model for hard diffraction , was also studied by the CDF Collaboration
at the Tevatron through diffractive W production
and dijet production \cite{cdfw}
\cite{cdfjet}, which give further evidence for the hard partonic stucture of
the Pomeron. Combining these two experiments, the CDF Collaboration
determined the hard-gluon content of the Pomeron to be $ f_g = 0.7 \pm 0.2$,
this result is also independent  of the Pomeron flux normalization or the
validity of the momentum sum rule for the Pomeron. 

\par
However, there is a problem. The I-S model is based on the assumption of
diffractive hard scattering factorization. Recently, a factorization theorem
has been proven by Collins \cite{jccollins} for the lepton induced
diffractive hard scattering processes,
such as DDIS and diffractive direct photoproduction of jets.
This factorization theorem justifies the analysis given by the ZEUS and H1
Collaboration, and establishes the the universality of the diffractive parton
distrubutions for those processes to which the theorem applies. In contrast,
no factorization theorem has been established for diffractive hard scattering
in hadron-hadron collisions. As already known before the advent of QCD,
factorization fails for hard processes in diffractive hadron-hadron
scattering \cite{p-qcd}. Furthermore, in order to preserve the shapes of
the $M^2$ and t distributions in soft single diffraction (SD) and predict
correctly the experimentally observed SD cross section at all energy in
$P{\bar P}$ collisions, Goulianos \cite{goulianos} has proposed to
renormalize the Pomeron flux in an energy-dependent way, this approach
indicates the breakdown of the triple-Regge theory for soft single
diffractive exciation, and implies that diffractive hard factorization
is likely breakdown in hadron-hadron collisions which appears to be
confirmed by CDF experiments\cite{cdfw}\cite{cdfjet}. At large $|t|$ , where
perturbative QCD applies to the Pomeron, it was proven that there is a leading
twist contribution which broken the factorization theorem for diffractive
hard scattering in hadron-hadron collision \cite{copom}. Evidence for this 
substantial coherent perturbative contribution has been observed by the 
UA8 experiment in diffractive jet production, in which the 
square of the proton's four-momentum transfer t is in the region $ -2 < t < 
-1 $ GeV$^2$ \cite{ua8}. On the other hand, for the Pomeron at small $|t|$
, nonperturbative QCD phycics dominates, where the 
I-S model without much of a coherent contribution would be appropriate.
Therefore, various hard diffractive processes may be considered to probe
the partonic struction of the Pomeron, and test the hard diffractive 
factorization in hadron-hadron collisions \cite{bergdf} \cite{testa}. 
Recently, Alero {\it et al.}\cite{testb} extract the parton densities of
the Pomeron from the HERA data on DDIS and diffractive
photoproduction of jets, and 
use the fitted parton densities to predict the diffractive production of
jets and weak bosons in $P{\bar P}$ collisions at the Tevatron. The predicted 
cross sections are substantially high than the experimental data, which 
signals a breakdown of hard scattering factorization in diffractive
hadron-hadron collisions.

\par
In this paper, we will discuss another diffractive process, the associated
$\jpsi + \gamma $ single diffractive production at large $P_T $:
\beq
P + {\bar P } \to P + \jpsi + \gamma + X    .
\eeq
This process is of special significance because the produced large $P_T $
$\jpsi $ is easy to detect through its leptonic decay modes and the
$\jpsi $'s large $P_T $ is balance by the associated high energy photon.
We will see through the following calculations, although the production
cross section of associated $\jpsi +
\gamma $ is sensitive to the color-octet matrix elements, the ratio
of the single diffractive production cross section to the inclusive
production cross section are not so. We find that the ratio is sensitive to
the product $D f_g$,
where D is the renormalization factor of the Pomeron flux(which indicates
the factorization broken effects,)and $f_g$ is
the hard gluon fraction in the Pomeron. The measurement of this process
at the Tevatron and LHC would shed light on the nature of the Pomeron
and test the diffractive hard scattering factorization.
Furthermore, this process is also interesting to the study of heavy
quarkonium production mechanism.

\par
Our paper is organized as follows. We describe in detail our calculation
scheme in Sec.II. A brief introduction of the heavy quarkonium production
mechanism, the hard subprocesses for associated $\jpsi + \gamma $
production and a summary of the kinematics related to the $\jpsi$ $P_T$
distribution are given in this section. Our results and discussions are
given in Sec. III.

\section{Calculating Scheme}
Based on the I-S model for diffractive hard scattering, the associated
$\jpsi + \gamma $ single diffractive production process at large $P_T$
consists of three steps(shown in Fig.1). First, the Pomeron is emitted
from the proton with a small squared
four-momentum transfer $|t|$ . Second, partons from the anti-proton and
the Pomeron scatter in the hard subprocesses that produce the almost
point-like large $P_T$ pair of $c{\bar c} $ and a associated photon.
In the third
step, $ \jpsi $ is produced from the point-like $c{\bar c}$ via
soft gluon radiation.

\subsection{Heavy quarkonium production mechnism}

Prior to 1993, the conventional wisdom for heavy quarkonium production
was based on the so-called color-singlet model (CSM)\cite{csm} which
assumes that the heavy quark pair is produced in a color-singlet state
with the right quantum numbers of the final heavy quarkonium  in the hard
subprocesses and the belief that the dominant production processes for
color-singlet $c{\bar c}$ were the Feynman diagrams that were lowest order
in $\alpha _s$ . However, at the Tevatron, the CSM prediction for prompt
$\jpsi$ production
is an order of magnitude smaller than the CDF data at large
$P_T$. The first major conceptual advance
in heavy quarkonium production
was the realization that fragmentation dominates at sufficiently large
$P_T$ \cite{fragm} which indicates that most charmonium at large $P_T$
is produced by the fragmentation of individual large $P_T$ partons.
The fragmentation functions are calculated in  CSM , for example, the
fragmentation fuction for $g\to \jpsi $ is calculated from the parton
process $ g\to c{\bar c} + gg $ in CSM . Including this fragmentation
mechanism brings the theoretical predictions for prompt $\jpsi$ production
at Tevatron to within a factor of 3 of the data \cite{bdfm}. But the
prediction for the $\psi^{\prime}$ remains a factor of 30 below the data
even after including the fragmentation contribution ( the $\psi^{\prime }$
``surplus'' problem ).
Furthermore, the presence of the logarithmic infrared divergences in
the production cross sctions for P-wave charmonium states and the
annihilation rate for $\chi_{cJ} \to q {\bar q} g $ indicate CSM
is incomplete. All these problems indicate that some important production
mechanism beyond CSM needs to be included \cite{pwcom}\cite{gcofrag}.
And so the color-octet mechanism is proposed which is based on the
factorization formalism of nonrelativistic quantum chromodynamics
(NRQCD)\cite{comfact}\cite{nrqcd}. Contrary to the basic assumption of CSM,
the heavy quark pair in a color-octet state can bind to
form heavy quarkonium.

\par
Although the color-octet fragmentation picture of heavy quarkonium production
\cite{gcofrag} has provided valuable insight, the approximation that enter
into fragmentation computations break down when a quarkonium's energy
becomes comparable to its mass. The fragmentation predictions for charmonium
production are therefore unrealiable at low $P_T$. In the case of
$\Upsilon $ production, the exist data at the Tevatron are in the
$P_T < 15$ GeV  region, which significantly disagree with the fragmentation
predictions. Based upon the above several recently developed ideas in
heavy quarkonium physics, Cho and Leibovich\cite{cho} identify a large class
of color-octet diagrams that mediate quarkonia production at all energies,
which reduce to the dominant set of gluon fragmentation graphs in the high
$P_T$ limit. By fitting the data of prompt $ \jpsi $ and $\Upsilon $
production at the Tevatron, numerical values for the long distance matrix
elements are extracted, which are generally consistent with NRQCD power
scaling rules \cite{power}.
In order to convincingly establish the color-octet
machanism, it is important to test whether the same matrix elements be able
to explain heavy quarkonim production in other high energy processes, such
as inclusive $\jpsi$ production in $e^+ e^-$ annihilation on the $Z^0$
resonace and inelastic $\jpsi $ photoproduction at HERA {\it et al. }.

\subsection{The hard subprocesses for associated $\jpsi + \gamma $
production}

$\jpsi $ is described within the NRQCD framework in terms of Fock state
decompositions as
\ba
|\jpsi> &=& O(1)~ |c{\bar c}[^3S_{1}^{(1)}]> +
            O(v) |c{\bar c}[^3P_{J}^{(8)}]g> + \nonumber \\
        & & O(v^2) |c{\bar c}[^1S_{0}^{(8)}]g> +
            O(v^2) |c{\bar c}[^3S_{1}^{(1,8)}]gg> + \nonumber \\
        & & O(v^2) |c{\bar c}[^3P_{J}^{(1,8)}]gg> + \cdots ,
\ea
where the $c{\bar c}$ pairs are indicated within the square brackets in
spectroscopic notation. The pairs' color states are indicate by singlet (1)
or octet (8) superscripts. The color octet $c{\bar c}$ state can make
a transition into a physical $\jpsi$ state by soft chromoelectric
dipole (E1) transition(s) or chromomagnitic M1 transition(s)
\beq
(c {\bar c})[^{2S + 1}L_{j}^{(1,8)}] \to \jpsi  .
\eeq
NRQCD factorization scheme \cite{comfact} has been established to
systematically separate high and low energy scale interactions.
It is based upon a bouble power series expasion in the strong interaction
fine structure constant $\alpha _s = \frac{g_s^2}{4 \pi}$ and the small
velocity parameter $v$. The production of a
$ (c {\bar c})[^{2S + 1}L_{j}^{(1,8)}] $ pair with separation less than
or of order $\frac{1}{m_c}$ are perturbatively computable. The long distance
effects for the produced almost point-like $c{\bar c}$ to form the bound
state are isolated into nonperturbative matrix elements. Furthermore,
 NRQCD power counting rules can be exploited to determine the dominant
contributions to various quarkonium processes\cite{power}.
For direct $\jpsi $ production, the color-octet matrix elements,
$\mtso , \moso , \mtpj $ are all scaling as $m_c^3 v_c^7 $. So these
color-octet contributions to $\jpsi $ production must be included for
consistency.

\par
The partonic level subprocesses for associated $\jpsi + \gamma $
production are composed of the gluon fusion subprocesses, which are
sketched in Fig.2.  These are
\ba
g + g & \to & \gamma + (c{\bar c})[^3S_{1}^{(1)}, ^3S_{1}^{(8)}]~,
 \nonumber \\
g + g & \to & \gamma + (c{\bar c})[^1S_{0}^{(8)}, ^3P_{J}^{(8)}]~ .
\ea

The quark initiated subprocesses($q{\bar q}$ channel) are strongly suppresed
and will be neglected further. The color-singlet gluon-gluon fusion
contribution to associated $\jpsi + \gamma $ production is well
known \cite{berg}:
\beq
\label{sg3s11}
\frac{d\hat{\sigma}} {d\hat{t}}(singlet) = \frac{{\cal N}_1}{16 \pi \hat{s}^2}
\Bigg[
\frac{\hat{s}^2 (\hat{s}-4m_c^2)^2 + \hat{t}^2 (\hat{t}-4m_c^2)^2 +
      \hat{u}^2 (\hat{u}-4m_c^2)^2 }
      { (\hat{s}-4m_c^2)^2 (\hat{t}-4m_c^2)^2 (\hat{u}-4m_c^2)^2 }
\Bigg],
\eeq
\noindent
where
\beq
\hat{s} = (p_1 + p_2)^2 ,~\hat{t} = (p_2 - P)^2 ,~\hat{u} = (p_1 - P)^2 .
\eeq

\noindent
The overall normalization ${\cal N}_1$ is defined as
\beq
{\cal N}_1 = \frac{4}{9} g_s^4 e^2 e_c^2 m_c^3 G_1(\jpsi),
\eeq
\noindent
where 
\beq
G_1(\jpsi) = \frac{\mtss }{3m_c^2} = \frac{\Gamma (\jpsi \to j^+ l^-)}
             {\frac{2}{3} \pi e_c^2 \alpha ^2 }
\eeq
and  $e_c^2 = \frac{2}{3}$ . \\
The average-squared amplitude of the subprocess
$ g + g  \to  \gamma + (c{\bar c})[^3S_{1}^{(8)}]$ can be obtained from the
average-squared amplitude of
$g + g  \to  \gamma + (c{\bar c})[^3S_{1}^{(1)}]$ by taking into account of
defferent  color factor. The result is
\ba
\label{sg3s18}
\frac{d\hat{\sigma}} {d\hat{t}}&~& [g + g  \to
   \gamma + (c{\bar c})[^3S_{1}^{(8)}]  \to \gamma + \jpsi] \nonumber \\
&=& \frac{1}{16 \pi \hat{s}^2 } \frac{15}{6} \overline{\Sigma }
    |M(g + g \to \gamma + (c{\bar c})[^3S_{1}^{(1)}] )|^2 \cdot
    \frac{1}{24 m_c} \mtso \nonumber \\
&=& \frac{1}{16 \pi \hat{s}^2 } \frac{15}{6} \frac{1}{8} \frac{1}{12} \cdot
  64 g_s^4 e_c^2 e^2 m_{\psi}^2
  \Bigg[
\frac{\hat{s}^2 (\hat{s}-4m_c^2)^2 + \hat{t}^2 (\hat{t}-4m_c^2)^2 +
      \hat{u}^2 (\hat{u}-4m_c^2)^2 }
      { (\hat{s}-4m_c^2)^2 (\hat{t}-4m_c^2)^2 (\hat{u}-4m_c^2)^2 }
\Bigg] \nonumber \\
&~& \cdot \frac{1}{24 m_c} \mtso .
\ea
The average-squared amplitudes of the subprocesses
$g + g  \to  \gamma + (c{\bar c})[^1S_{0}^{(8)}]$ and
$g + g  \to  \gamma + (c{\bar c})[^3P_{J}^{(8)}]$ can be found in \cite{kim},

\ba
\label{sgsp08}
\frac{d\hat{\sigma}} {d\hat{t}}&~& [g + g  \to
   \gamma + (c{\bar c})[^1S_{0}^{(8)}]  \to \gamma + \jpsi] \nonumber \\
&=& \frac{1}{16 \pi \hat{s}^2 } \overline{\Sigma }
    |M(g + g \to \gamma + (c{\bar c})[^1S_{0}^{(8)}] )|^2 \cdot
    \frac{1}{8 m_c} \moso , \nonumber \\ 
\frac{d\hat{\sigma}} {d\hat{t}}&~& [g + g  \to
   \gamma + (c{\bar c})[^3P_{J}^{(8)}]  \to \gamma + \jpsi] \nonumber \\
&=& \frac{1}{16 \pi \hat{s}^2 } \sum_{J} \overline{\Sigma }
    |M(g + g \to \gamma + (c{\bar c})[^3P_{J}^{(8)}] )|^2 \cdot
    \frac{1}{8 m_c} \mtpo ,
\ea
where the heavy quark spin symmetry
\beq
\mtpj = (2 J + 1 ) \mtpo
\eeq
is exploited.

\subsection{The $P_T$ distribution of $\jpsi $ }

Now we consider the $P_T $ distribution of $\jpsi $ produced in process
\beq
\label{proc}
p(P_p) + {\bar p}(P_{\bar p}) \to p(P^{\prime}_{p}) + \jpsi (P) + \gamma (k)
 + X .
\eeq
Based on the I-S model for diffractive hard scattering, the differential
cross section can be expressed as

\beq
d\sigma = f_{I\!P/p }(\xi,t) f_{g/I\!P} (x_1, Q^2) f_{g/{\bar p}} (x_2, Q^2)
          \frac{d\hat{\sigma}}{d\hat{t}} d \xi dt d x_1 dx_2 d \hat{t} ,
\eeq
where $\xi $ is the momentum fraction of the proton carried by the Pomeron,
$t = ( P_p - P_{p}^{\prime} )^2 $ is the squared of the proton's
four-momentum transfer. $f_{I\!P/p }(\xi,t)$ is the Pomeron flux factor 
\ba
f_{I\!P/p }(\xi,t)& = & \frac{d^2 \sigma_{SD} /d\xi d t }{\sigma_{T}^{
      I\!P P} (s^{\prime}, t) } = \frac{\beta_1^2 (0) }{16 \pi} \xi ^{
      1 - 2 \alpha (t) } F^2(t) \nonumber \\
& = & K \xi ^{1 - 2 \alpha (t) } F^2(t)  ,
\ea
where the parameters are choosen as \cite{goulianos}
\beq
K = 0.73 GeV^2 , ~\alpha (t) = 1 + 0.115 + 0.26 (GeV^{-2}) {\large t} ,
~F^2(t) = e^{4.6 t} .
\eeq
In the following calculation, we use the renormalized flux factor for the
Pomeron, proposed by Goulianos \cite{goulianos} in order to preserve
the shapes of the $M^2$ and $t$ distribution in soft single diffraction (SD)
and predict correctly the observed SD cross section at all energies in $
p{\bar p}$ collisions,

\beq
f^{RN}_{I\!P/p }(\xi,t) = D f_{I\!P/p }(\xi,t) ,
\eeq
the renormalization factor $ D $ is defined as
\beq
\label{dfactor}
D = Min(1, \frac{1}{N})
\eeq
with
\beq
\label{nv}
N = \int_{\xi_{min}}^{\xi_{max}} d \xi \int^{0}_{- \infty} d t
f_{I\!P/p }(\xi,t) ,
\eeq
where $\xi_{min} = \frac{M_0^2}{s} $ with $M_0^2 = 1.5 GeV^2 $(effective
threshold) and $\xi_{max} = 0.1 $(coherence limit). At the Tevatron energy
($ \sqrt{s} = 1800 GeV $), $ D = \frac{1}{9} $.

\par
We now consider the kinematics. In the $p{\bar p}$ c.m. frame, we can express
the momenta of the incident $p, {\bar p}$ and gluons {\it etc.} as
\ba
P_p        &=& \frac{\sqrt{s}}{2} (1, ~1, ~\vec{0} ) \nonumber \\
P_{\bar p} &=& \frac{\sqrt{s}}{2} (1, ~- 1, ~\vec{0} ) \nonumber \\
P_{I\!P}   &=& \frac{\sqrt{s}}{2} \xi (1, ~1, ~\vec{0} ) \nonumber \\
p_1        &=& \frac{\sqrt{s}}{2} x_1 \xi (1, ~1, ~\vec{0} )  = x_1 P_{I\!P}
           \nonumber \\
p_2        &=& \frac{\sqrt{s}}{2} x_2  (1, ~- 1, ~\vec{0} )  = x_2 P_{\bar p}
\ea
where the first component is the energy, the second is longitudinal
momentum, and the third is the transverse component of the four-momentum,
$x_1, x_2 $ are the momentum fractions of the gluons. The momenta of the
outgoing $\jpsi $ are given by
\beq
P = (E, P_L, P_T) = (E , P_T sh \eta ,\vec{P}_T ) ,
\eeq
where $P_T$ is the transverse momentum of  $\jpsi$, $\eta$ is the
pseudo-rapidity of $\jpsi$ and $ E = \sqrt{m_{\psi}^2 + P_T^2 ch ^2 \eta} $ .

\par
The Mandelstam variables are defined as
\ba
\label{hatstu}
s       &=& (P_p + P_{\bar p})^2 , \nonumber \\
\hat{s} &=& (p_1 + p_2)^2 = x_1 x_2 \xi s , \nonumber \\
\hat{t} &=& (p_2 - P)^2 = m_{\psi}^{2} - \sqrt{s} x_2 ( E + P_T sh \eta )
           , \nonumber \\
\hat{u} &=& (p_1 - P)^2 = m_{\psi}^{2} - \sqrt{s} x_1 \xi ( E - P_T sh \eta ) .
\ea
Using$ \hat{s} + \hat{t} + \hat{u} = m_{\psi}^2 $, we have
\beq
\label{eqx2}
x_2 = \frac{\sqrt{s} \xi x_1 (E - P_T sh \eta) - m_{\psi}^2 }
           {x_1 \xi s - \sqrt{s} (E - P_T sh \eta)}
\eeq
In order to obtain the distribution in the transverse momentum $P_T$ for
the process Eq.(\ref{proc}), we express the differential cross section as

\beq
d\sigma = f_{I\!P/p }^{RN} (\xi,t) f_{g/I\!P} (x_1, Q^2) f_{g/{\bar p}}
          (x_2, Q^2) \frac{d\hat{\sigma}}{d\hat{t}}
          J(\frac{x_1 x_2 \hat{t}}{x_1 \eta P_T})
          d \xi dt d x_1 d \eta d P_T ,
\eeq
where the Jacobian can obtain from Eqs.(\ref{hatstu}) and Eq.(\ref{eqx2}),
\beq
J(\frac{x_1 x_2 \hat{t}}{x_1 \eta P_T})
= \frac { 2 s x_1 x_2 \xi P_T^2 ch \eta }
        {E [ x_1 \xi s - \sqrt{s} (E + P_T sh \eta)]} .
\eeq
Then the $P_T$ distribution of $\jpsi$ is expressed as
\beq
\label{ptdis}
\frac{d\sigma }{d P_T}
= \int_{\eta_{min}}^{\eta_{ma}} d \eta \int_{\xi_{dw}}^{
  \xi_{up}} d \xi \int_{x_{1 min}}^{x_{1 max}} d x_1 \int_{-1}^{0} d t
f_{I\!P/p }^{RN} (\xi,t) f_{g/I\!P} (x_1, Q^2) f_{g/{\bar p}}
          (x_2, Q^2) J(\frac{x_1 x_2 \hat{t} }{x_1 \eta P_T} )
         \frac{d\hat{\sigma}}{d\hat{t}} ,
\eeq
where the allowed regions of $ x_1, \xi $ are given by
\ba
x_{1 min } &=& \frac{ E + P_T sh \eta - \frac{m_{\psi}^2 }{\sqrt{s} } }
                    { \xi ( \sqrt{s} - E + P_T sh \eta)} , \nonumber \\
\xi_{dw}   &=& \frac{ E + P_T sh \eta - \frac{m_{\psi}^2 }{\sqrt{s} } }
                    {  \sqrt{s} - E + P_T sh \eta } .
\ea
In order to suppress the Reggon contributions, we set $\xi_{up} = 0.05$ as
usual .

\section{Numerical results and discussions}
Now, we are ready to show the numberical results from the analytic
expressions giving in the previous section. For numberical predictions
we use $m_c = 1.5 GeV, \Lambda_4 = 235 MeV $, and set the factoriztion
scale and the renormalization scale both equal to the transverse mass
of $\jpsi$, {\it i.e.}, $Q^2 = m_T^2 = (m_{\psi}^2 + P_T^2)$ .
For the color-octet matrix elements $\mtso , \moso$ and $\mtpo $ we
use the values determined by Beneke and Kr$\ddot{a}$mer \cite{beneke}
from fitting the direct $\jpsi$ production data at
the Tevatron \cite{direct} using GRV LO parton
distribution functions,
$$ \mtso = 1.12 \times 10^{-2} GeV^3 ,$$  
\beq
\label{matrix}
\moso + \frac{3.5}{m_c^2} \mtpo = 3.90 \times 10^{-2} GeV^3 .
\eeq
Since the matrix elements $\moso $ and $\mtpo$ are not determined separately,
we present the two extreme values allowed by Eq.(\ref{matrix}) as
\ba
^1S_0 ~saturated ~case :~~ &\moso &= 3.90 \times 10^{-2} GeV^3  \nonumber \\
~                        &\mtpo &= 0 ,   \nonumber \\
^3P_J ~saturated ~case :~~ &\mtpo &= 1.11 \times 10^{-2} mc^2 GeV^3  \nonumber \\
~                        &\moso &= 0 .
\ea
For the parton distribution functions, we use the GRV LO gluon
distribution function for the anti-proton \cite{grv} and the hard gluon
distribution fundtion for the Pomeron \cite{cdfjet}
\beq
x f_{g/I\!P} (x, Q^2) = f_g 6 x (1-x) , ~f_g = 0.7 \pm 0.2 . 
\eeq
We use the central value of $f_g$ above for numberical calculation. We
neglect any $Q^2$ evolution of the gluon density of the Pomeron at present
 stage.

\par
With all ingredients set as above ,
in Fig.3 we show the $P_T$ distribution
$ B \frac{d \sigma} {d P_T} $ for single diffractive $\jpsi + \gamma$
production in $p{\bar p}$ collisions at $\sqrt{s} = 1.8 $TeV , integrated
over a pseudo-rapidity region $-1 \leq \eta \leq 1$ (central region).
Where $ B = 0.0594$ is the $\jpsi \to \mu^+ \mu^- $ leptonic
decay branching ratio.
The lower solid line is the color-singlet gluon-gluon fusion contribution,
the lower dashed line represents $ ^1 S_{0}^{(8)}$-saturated color-octet
contribution, the lower dotted one is $^3P_{J}^{(8)}$-saturated color-octet
contribution and the lower dash-dot-dotted line is $^3S_{1}^{(8)}$
color-octet contribution.  For comparision, we also calculate the inclusive
associated $\jpsi + \gamma$ production $ p + {\bar p} \to
\jpsi + \gamma + X $ in the same kinematic region, the results are shown
as the upper lines of Fig.3. The code for the lines are the same as the
single diffractive production case. As shown in Fig.3 , 
the color-octet $^3S_{1}^{(8)}$ contribution is strongly suppressed
compared with the others over the entire $P_T$ region considered in both
production cases. The $^1S_{0}^{(8)}$ -saturated and
$^3P_{J}^{(8)}$-saturated contribution are smaller than the singlet
contribution  where $P_T < 5$ GeV,  though their contribution dominate in
the high $P_T$ region, their
differential cross section in the high
$P_T$ region is much small than that at low $P_T$ region , so integrated
over the $P_T$ region, their contribution ( $ B \sigma = 0.08 $ pb and $16$
 pb for $^1 S_{0}^{(8)}$-saturated contribution to SD and
inclusive production  respectively;
 $ B\sigma = 0.06$ pb and $12$ pb  for
$^3P_{J}^{(8)}$-saturated contribution to SD  and inclusive production
respectively ) are smaller than
the color-singlet  contributoin ($ B\sigma (singlet) = 0.10 $ pb and $20$ pb
for SD and inclusive production respectively ).
The total SD  (inclusive)
production cross section at $\sqrt{s} = 1.8$ TeV, intergrated over
$4\leq P_T \leq 10 $ Gev  in central region is
$ B \sigma^{SD}(cen.) = 0.18 $ pb
 ( $ B \sigma^{Inclusive} (cen.) = 36$ pb  ) for
$^1 S_{0}^{(8)}$-saturated case, $ B \sigma^{SD}(cen.) = 0.16 $pb 
( $ B \sigma^{Inclusive}(cen.) = 32 $ pb )
for $^3P_{J}^{(8)}$-saturated case.
The ratio of the total SD production cross section to that of inclusive
production  in the central region $ R^{cen.} $ are $ 0.50 \% $ for both
$^1 S_{0}^{(8)}$-saturated case and $^3P_{J}^{(8)}$-saturated case.

\par
In Fig.4 , we show the $P_T$ distribution of $\jpsi$ integrated over
a pseudo-rapidity region$-4\leq \eta \leq -2 $ (forward region),
at $\sqrt{s}=1.8 $TeV, the code of lines is the same as Fig.3. One
can find the differential cross section is significantly smaller than
that in central region. So the diffractively produced $\jpsi $ should
be concentrated in the central detectors of collider. The same character
was found early by Berger {\it et al.} \cite{bergdf} in the rapidity
distribution of SD production of bottom and top quarks.
This contrasts with
the naive expectation that diffractively producted system appera only
at large rapidity.
The total SD  (inclusive)
production cross section at $\sqrt{s} = 1.8$ TeV, intergrated over
$4\leq P_T \leq 10 $Gev in forward region is
$ B \sigma^{SD}(fwd.) = 5.7 \times 10^{-2} $pb 
( $ B \sigma^{Inclusive} (fwd.) = 23$ pb  ) for
$^1 S_{0}^{(8)}$-saturated case, $ B \sigma^{SD} (fwd.)  = 5.1 \times 10^{-2}$
pb ( $ B \sigma^{Inclusive} (fwd.) = 20 $ pb ) for $^3P_{J}^{(8)}$-saturated
case. The ratio of the total SD production cross section to that of inclusive
production  in the forward region $R^{fwd.}$  are $ 0.25 \% $ for both
$^1 S_{0}^{(8)}$-saturated case and $^3P_{J}^{(8)}$-saturated case.

\par
In TABLE I. , we show the ratio
\beq
R(P_T) = \frac{ d \sigma ^{SD} }{d P_T} / \frac{d \sigma ^{Inclusive} }
              {d P_T }
\eeq
at $\sqrt{s} = 1.8 $ TeV in central region and forward region, where
\beq
\frac{d \sigma ^{SD} }{d P_T} =
\frac{d \sigma ^{SD}(singlet)} {d P_T} +
\frac{d \sigma ^{SD}(^3S_{1}^{(8)}) }{d P_T} +
\frac{d \sigma ^{SD}(^1S_{0}^{(8)}) }{d P_T}
\eeq
for $^1 S_{0}^{(8)}$-saturated case, and
\beq
\frac{d \sigma ^{SD} }{d P_T} =
\frac{d \sigma ^{SD}(singlet)}{d P_T} +
\frac{d \sigma ^{SD}(^3S_{1}^{(8)}) }{d P_T} +
\frac{d \sigma ^{SD}(^3P_{J}^{(8)}) }{d P_T}
\eeq
for $^3P_{J}^{(8)}$-saturated case.

\begin{table}[tbp]
\caption{ The ratio $R(P_T)$ as a function of $P_T$ at
the Tevatron
in central region $(-1 \leq \eta \leq 1) ~R^{cen.} (P_T)$  and
forward region $(-4 \leq \eta \leq -2) ~R^{fwd.} (P_T)$. }
\begin{center}
\begin{tabular}{|c|c|c|c|c|c|c|c|}
\hline
$P_T$ (GeV)          &  4   &  5   &  6   &  7   &  8   &    9 & 10  \\
\hline
$R^{cen.} (P_T) (\%)$& 0.52 & 0.52 & 0.50 & 0.48 & 0.47 & 0.46 & 0.44 \\
\hline
$R^{fwd.} (P_T) (\%)$& 0.28 & 0.28 & 0.28 & 0.28 & 0.29 & 0.29 & 0.30 \\
\hline
\end{tabular}
\end{center}
\end{table}

The ratio $ R(P_T)$ for $^1 S_{0}^{(8)}$-saturated case
is the same as that for $^3P_{J}^{(8)}$-saturated case.
From this table, we can see that $R(P_T)$ is almost constant for
$P_T \leq 6 $ GeV  both in central and forward regions. $R(P_T)$
varies slowly as $P_T$ increase. Because  
the differential cross section in the high
$P_T$ region is much small than that at low $P_T$ region , integrated
over the overall $P_T$ region, the ratios $R^{cen.} $ and $R^{fwd.} $
are not sensitive to the $P_T$ smearing effects . Forthermore , we
have varied the color-octet matrix elements $\mtso, \moso + \frac{3.5}{m_c^2}
\mtpo $ by multiplied them by a factor between $ \frac{1}{10}$ and $ 2$,
the result ratios $R^{cen.} $ and $R^{fwd.} $ is the same. This character
demonstrates that the ratios $R^{cen.} $ and $R^{fwd.} $ are not
sensitive to the values of color-octet matrix elements. The
ratios $R^{cen.} $ and $R^{fwd.} $ are
proportional to $D f_g$ , hence
are sensitive to the gluon fraction of the Pomeron $f_g$ and the
renormalization factor $ D $ which indicates the factorization  broken
effects. From other single diffractive production experiments, $f_g$ can
be determined, as the CDF Collaboration at Tevotran and the ZEUS
Collaboration at HERA DESY have done, the renormalization factor $D$ can
be determined precisely from those ratios, {\it vice versa}. So measuring
these ratios can probe the gluon density in the Pomeron and shed light
on the test of diffractive hard scattering factoriztion theorem.

\par
In Fig.5, we show the $P_T$ distribution of $\jpsi$ integrated over
a pseudo-rapidity region$ -1 \leq \eta \leq 1 $ ( central ) 
at LHC energy $\sqrt{s} = 14 $TeV with the factor $ D = 0.052 $ calculated
from Eq.(\ref{dfactor}). The code of lines is the same as Fig.3.
As expected, the increasing of the
differential cross section with c.m energy is slowed down by the
renormalization of the Pomeron flux.
The total SD  (inclusive)
production cross section at $\sqrt{s} = 14$ TeV, intergrated over
$4\leq P_T \leq 10 $Gev in central region is
$ B \sigma^{SD}(cen.) = 0.51 $pb 
( $ B \sigma^{Inclusive} (cen.) = 3.4 \times 10^2$ pb  ) for
$^1 S_{0}^{(8)}$-saturated case, $ B \sigma^{SD} (cen.)  = 0.46 $
pb ( $ B \sigma^{Inclusive} (cen.) = 3.0 \times 10^2 $ pb ) for
$^3P_{J}^{(8)}$-saturated case.
The ratio of the total SD production cross section to that of inclusive
production  in the central region $R^{cen.}$  are $ 0.15 \% $  for both
$^1 S_{0}^{(8)}$-saturated case and $^3P_{J}^{(8)}$-saturated case.

\par
In the above calculations, we have set the values of factor $D$
according to Eq.(\ref{dfactor}) , it is $\frac{1}{9}$ at Tevatron
energy and $0.052 $ at LHC energy. But the value is not unique, since
it may change with different choice of the parameters such as $M_0$ and
$\xi_{max}$ in Eq.(\ref{nv}). If we use the central value of
$D = 0.18 \pm 0.04  $ measured by the CDF Collaboration at
the Tevatron, the predicted ratio in the forward region $R^{fwd.}$
at the Tevatron , taking into account both$ p, {\bar p}$ can be diffractively
scattered, will be $ 0.81\%$ , which is close to the measured rate of
diffractive dijet production with two jets of $E_T > 20 $GeV , $1.8 <
|\eta| < 3.5 $ and $\eta_1 \eta_2 > 0$  .
\par
Experimentally, the nondiffractive background to the diffractive associated
$\jpsi + \gamma$ production must be dropped out in order to obtain useful
information about the natrue of Pomeron and the factorization broken effects,
this can be attained by performing the rapid gap enalysid ad in the CDF
diffractive dijet experiment.

\par
In conclusion, in this paper we have shown that the diffractive
associated $\jpsi + \gamma $ production at large $P_T$ is sensitive
to the gluon content of the Pomeron and the factorization broken
effects in hard diffractive scattering. Though the sigle diffractive
and inclusive production  cross section is sensitive to the values of
coler-octet matrix elements, the ratio of single diffractive to inclusive
$\jpsi + \gamma $ production is not so and
proportional to $D f_g$ , hence
are sensitive to the gluon fraction of the Pomeron $f_g$ and the
renormalization factor $ D $ which indicates the factorization  broken
effects. So experimental measurement of
this ratio at the Tevatron and LHC can shed light on the nature of Pomeron
and test the assumption of diffractive hard scattering fratorization.

\vskip 1cm
\begin{center}
{\bf\large Acknowledgments}
\end{center}

We thank Professor Kuang-Ta Chao and Dr. Feng Yuan for their reading of
the manuscript and useful discussions.
This work is supported in part by the National Natural Science Foundation of
China, Doctoral Program Foundation of Institution of Higher Education of China
and Hebei Natural Province Science Foundation, China.


\newpage
\centerline{\bf \large Figure Captions}
\vskip 2cm
\noindent
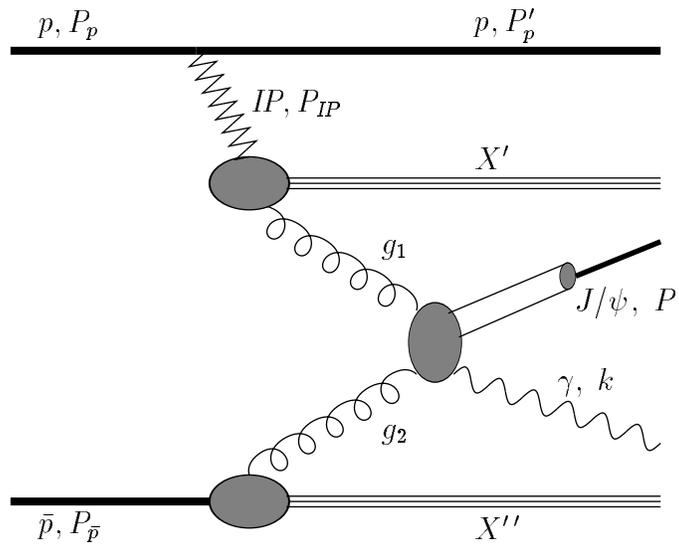
Fig.1. Sketch diagram for diffractive associated $\jpsi + \gamma $
production in the Ingelman-Schlein model for diffractive hard scattering.

\noindent
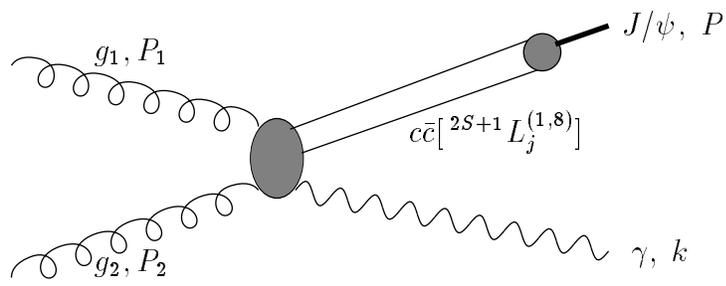
Fig.2. The subprocess $ g_1 + g_2 \to \gamma + c{\bar c}[^{2S + 1}L^{(1,8)}
_{J}] \to \gamma + \jpsi$ in the NRQCD framework.

\noindent
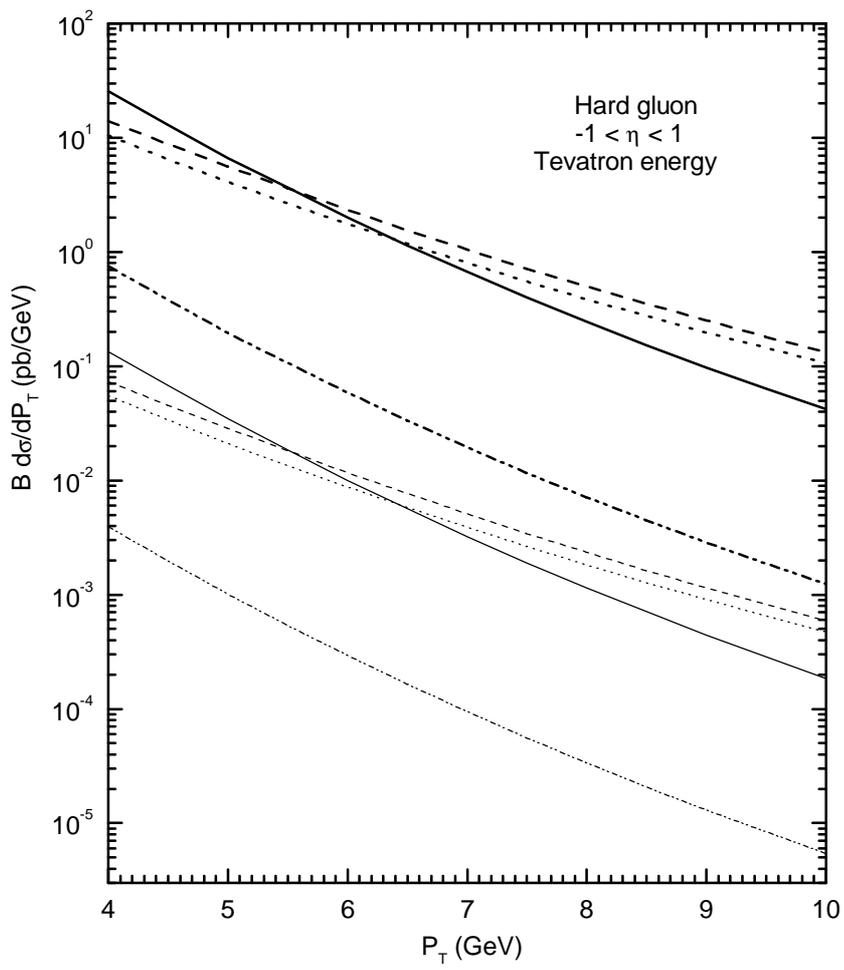
Fig.3. Transverse momentum of $\jpsi (P_T) $ distribution$ B d\sigma /dP_T $
, integrated over the $\jpsi$ pseudo-rapidity range $|\eta| \leq 1 $
(central region), for single diffractive (lower) and inclusive (upper)
associated $\jpsi + \gamma $ production at the Tevatron. Here $B$ is
the branching ratio of $\jpsi \to \mu^+ \mu^-  (B = 0.0594)$.
The solid line is the color-singlet gluon-gluon fusion contribution,
the dashed line represents $ ^1 S_{0}^{(8)}$-saturated color-octet
contribution, the dotted one is $^3P_{J}^{(8)}$-saturated color-octet
contribution and the dot-dot-dashed line is $^3S_{1}^{(8)}$
color-octet contribution.

\noindent
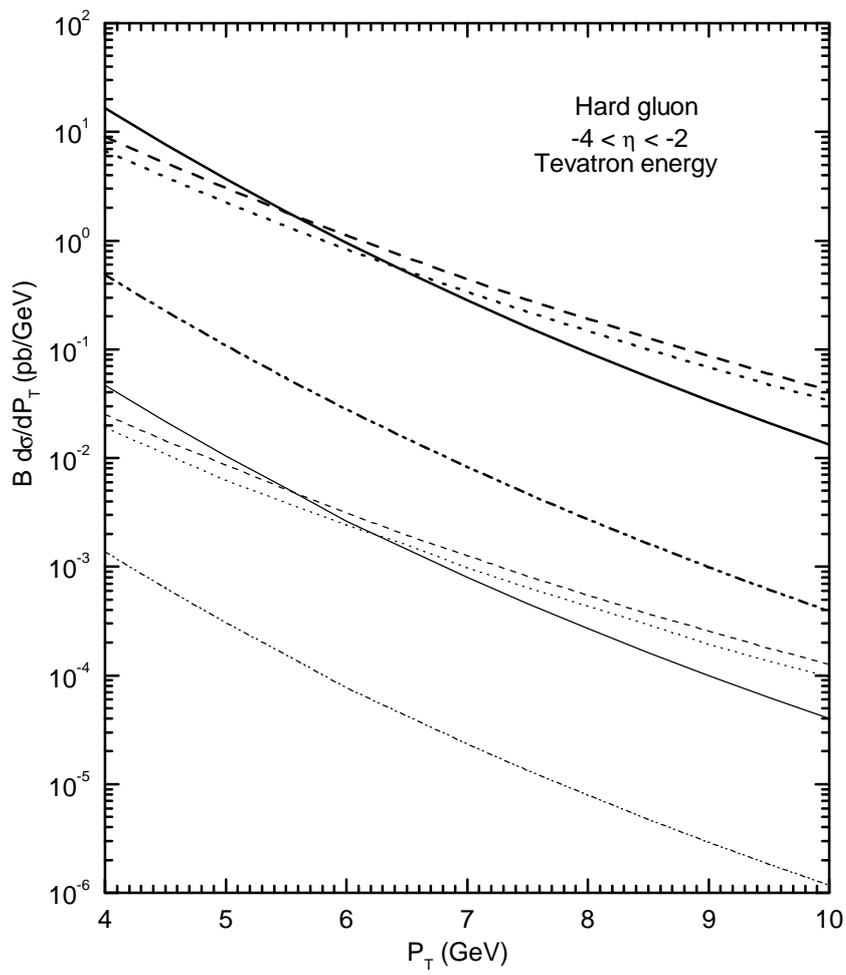
Fig.4. Transverse momentum of $\jpsi (P_T) $ distribution$ B d\sigma /dP_T $
, integrated over the $\jpsi$ pseudo-rapidity range $-4 \leq \eta \leq -2 $
(forward region), for single diffractive (lower) and inclusive (upper)
associated $\jpsi + \gamma $ production at the Tevatron. Here $B$ is
the branching ratio of $\jpsi \to \mu^+ \mu^-  (B = 0.0594)$.
The solid line is the color-singlet gluon-gluon fusion contribution,
the dashed line represents $ ^1 S_{0}^{(8)}$-saturated color-octet
contribution, the dotted one is $^3P_{J}^{(8)}$-saturated color-octet
contribution and the dot-dot-dashed line is $^3S_{1}^{(8)}$
color-octet contribution.

\noindent
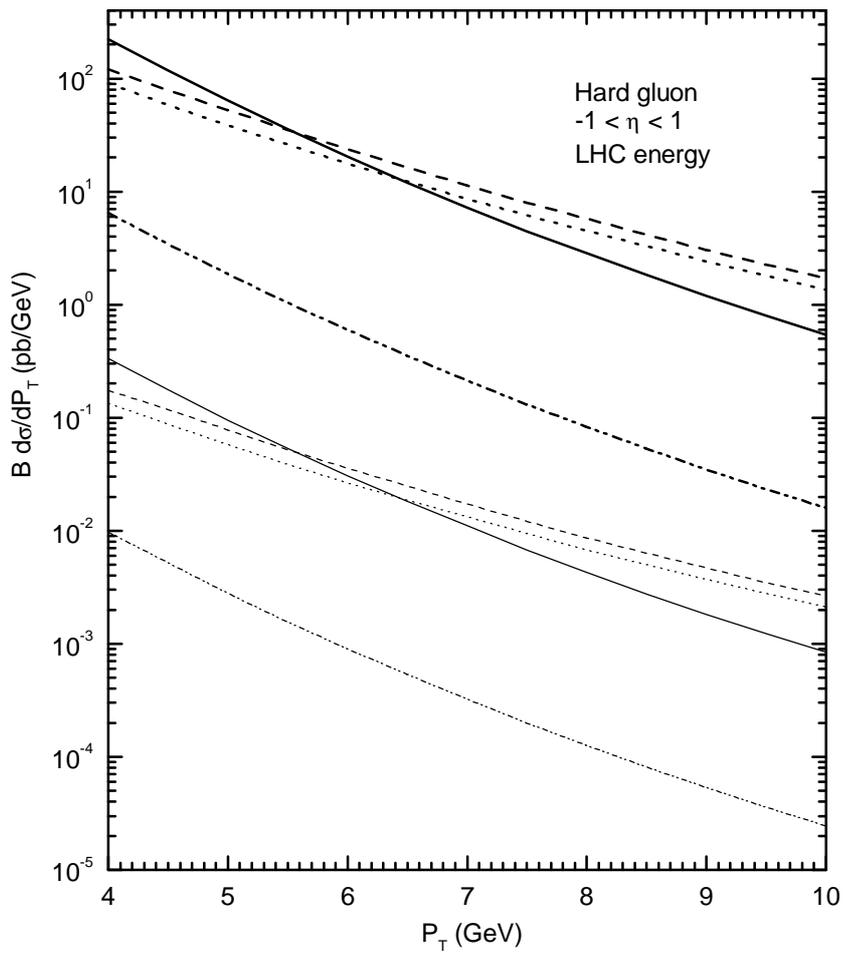
Fig.5. Transverse momentum of $\jpsi (P_T) $ distribution$ B d\sigma /dP_T $
, integrated over the $\jpsi$ pseudo-rapidity range $|\eta| \leq 1 $
(central region), for single diffractive (lower) and inclusive (upper)
associated $\jpsi + \gamma $ production at LHC. Here $B$ is
the branching ratio of $\jpsi \to \mu^+ \mu^-  (B = 0.0594)$.
The solid line is the color-singlet gluon-gluon fusion contribution,
the dashed line represents $ ^1 S_{0}^{(8)}$-saturated color-octet
contribution, the dotted one is $^3P_{J}^{(8)}$-saturated color-octet
contribution and the dot-dot-dashed line is $^3S_{1}^{(8)}$
color-octet contribution.


\begin{thebibliography}{99}
\bibitem{collins} P.D.B. Collins, {\it An introduction to Regge theory and
High Energy Phycics } ( Cambridge University Press, Cambridge, England,
1977 ). \\
K. Goulianos, Phys. Rep. {\bf 101}, 169 (1985).

\bibitem{low} S.Nussinov, Phys. Rev. Lett. {\bf 34}, 1286 (1975) ;
              Phys. Rev. {\bf D 14}, 246 (1976)  \\
              F. E. Low , Phys. Rev. {\bf D 12}, 163 (1975) 

\bibitem{gap} M. Derrick {\it et al.}, ZEUS Collaboration, Phys. Lett.
             {\bf B 315}, 481 (1993); {\it ibid.} {\bf B332},228 (1994);
T.Ahmed {\it et al.}, H1 Collaboration, Nucl. Phys. {\bf B 429}, 477 (1994)

\bibitem{ingelmam} G. Ingelman and P.E. Schlein , Phys. Lett. {\bf B 152},
    256 (1985) 

\bibitem{ua8} A. Brandt {\it et al.}, UA8 Collaboration, Phys. Lett.
{\bf B 297}, 417 (1992) ; R. Bonino {\it et al.}, {\it ibid.} {\bf B 211},
239 (1988) 

\bibitem{zeus} M. Derrick {\it et al.}, ZEUS Collaboration, Z. Phys.
{\bf C 68}, 569 (1995) ; M. Derrick {\it et al.}, ZEUS Collaboration,
Phys. Lett. {\bf B 356}, 129 (1995)

\bibitem{h1} T. Ahmed {\it et al.}, H1 Collaboration,  Phys. Lett.
{\bf B 348}, 681 (1995) ; Z. Phys. {\bf C 76}, 613 (1997) ( hep-ex/9708016 )

\bibitem{cdfw} F. Abe {\it et al.}, CDF Collaboration, Phys. Rev. Lett.
{\bf 78}, 2698 (1997)

\bibitem{cdfjet} F. Abe {\it et al.}, CDF Collaboration, Phys. Rev. Lett.
{\bf 79}, 2636 (1997)

\bibitem{jccollins} J.C. Collins, Phys. Rev. {\bf D 57}, 3051 (1998)

\bibitem{p-qcd}P.V. Landshoff and J.C. Polkinghorne , Nucl. Phys.
{\bf B 33}, 221 (1971) ;
F. Henyey and R. Savit, Phys. Lett. {\bf B 52}, 71 (1974) ;
J.L. Cardy and G.A. Winbow, Phys. Lett. {\bf B 52}, 95 (1974) ;
C.E. De Tar , S.D. Ellis and P.V. Landshoff, Nucl. Phys. {\bf B 87},
176 (1975)

\bibitem{goulianos} K. Goulianos , Phys. Lett. {\bf B 358}, 379 (1995);
 Erratum: {\it ibid.} {\bf B 363}, 268 (1995) ;
 K. Goulianos and J. Montanha, hep-ph/9805496


\bibitem{copom} J.C. Collins, L. Frankfurt and M. Strikmam,
Phys. Lett. {\bf B 307}, 161 (1993)

\bibitem{bergdf} E.L. Berger, J.C. Collins, D.E. Soper and
G. Stermam, Nucl. Phys. {\bf B 286},704 (1987)

\bibitem{testa} 
P. Bruni and G. Ingelman, Phys. Lett. {\bf B 311}, 317 (1993) ; \\
J.C. Collins {\it et al.}, Phys. Rev. {\bf D 51}, 318 (1995) ;\\
Hong-An Peng, Ke-Cheng Qin and Zhen-Min He ,
High Energy Phys. and Nucl. Phys.(in Chinese) , {\bf 18}, No.12, 1078 , 
{\it ibid.} (in English),  {\bf 19}, No.1, 31 (1995) ;\\
Hong-An Peng, Ke-Cheng Qin and Zhen-Min He ,
High Energy Phys. and Nucl. Phys.(in Chinese), {\bf 19}, No.1, 34 (1995),
{\it ibid.} (in English),  {\bf 19}, No.1, 51 (1995) ; \\
Feng Yuan and Kuang-Ta Chao, Phys. Rev. {\bf D 57}, 5658 (1998) 

\bibitem{testb} L. Alvero, J.C. Collins, J. Terron, and
J.J. Whitmore, hep-ph/9805268;
L. Alvero, J.C. Collins and J.J. Whitmore, hep-ph/9806340 

\bibitem{csm} Chao-Hsi Chang, Nucl. Phys. {\bf B 172}, 425 (1980) ;
        E. L. Berger and D. Jones: Phys. Rev. {\bf D 23}, 1512 (1981) ;
        R. Baier and R. Rukel: Z. Phys. {\bf C19} 251 (1983) 

\bibitem{ua1} C. Albajar {\it et al.}, UA1 Collaboration,
   Phys. Lett. {\bf B 256}, 112 (1991) 

\bibitem{fragm} E. Braaten and T.C. Yuan, Phys. Rev. Lett. {\bf 71},
 1673 (1993)

\bibitem{bdfm} E. Braaten, M.A. Doncheski, S. Fleming and M. Mangano,
  Phys. Lett. {\bf B 333}, 548 (1994) ;
  M. Cacciari and M. Greco, Phys. Rev. Lett. {\bf 73}, 1586 (1994) ;
  D.P. Roy and K. Sridhar, Phys. Lett. {\bf B 339}, 141 (1994)
  
\bibitem{pwcom} G.T. Bodwin, E. Braaten and G.P. Lepage,
Phys. Rev. {\bf D 46}, R1914 (1992); {\it ibid.} {\bf D 46}, R3703 (1992);

\bibitem{gcofrag} E. Braaten and S. Fleming,
Phys. Rev. Lett. {\bf 74}, 3327 (1995);

\bibitem{review} For a recent review, see
E. Braaten, S. Fleming and T.C. Yuan, Annu. Rev. Nucl. Part. Sci. {\bf 46},
197 (1996); hep-ph/9602374

\bibitem{comfact} G.T. Bodwin, E. Braaten and G.P. Lepage,
Phys. Rev. {\bf D 51}, 1125 (1995); Eraatum:{\it ibid.} {\bf D 55},
5853 (1997)

\bibitem{nrqcd} W.E. Caswell and G.P. Lepage, Phys. Lett. {\bf B167},
    437 (1986)

\bibitem{cho} P. Cho and A.K. Leibovich, Phys. Rev. {\bf D 53},150 (1996) ;
  {\it ibid.} {\bf D 53}, 6203 (1996) 

\bibitem{power} G.P. Lepage, L. Magnea, C. Nakhleh, U. Magnea and
K. Hornbostel, Phys. Rev. {\bf D 46}, 4052 (1992)

\bibitem{berg} E. L. Berger and D. Jones: Phys. Rev. {\bf D 23},
1512 (1981) ;

\bibitem{kim} C.S. Kim, Jungil Lee and H.S. Song, Phys. Rev. {\bf D 55}, 5429
(1997); P. Ko, Jungil Lee and H.S. Song, Phys. Rev. {\bf D 54}, 4312 (1996)

\bibitem{beneke} M. Beneke and M. Kr$\ddot{a}$mer, Phys. Rev. {\bf D 55},
R5269 (1997)

\bibitem{direct}F. Abe {\it et al.}, CDF Collaboration, Phys. Rev. Lett.
 {\bf 79},578 (1997)

\bibitem{grv} M. Gl$\ddot{u}$ck, E. Reya and A. Vogt, Z. Phys. {\bf C 67},
433 (1995) 
\end{thebibliography}
\end{document}